\documentstyle [12pt] {article}
\title{A CLASSICAL AND QUANTUM MECHANICAL ANALOG OF TWO CAPACITORS PARADOX}

\author{Vladan Pankovi\'c, Darko V. Kapor \\
Department of Physics, Faculty of Sciences, 21000 Novi Sad,\\ Trg
Dositeja Obradovi\'ca 4., Serbia, vpankovic@if.ns.ac.yu}

\date {}
\begin{document}
\maketitle \vspace {0.5cm}
 \vspace {0.1cm}

\begin {abstract}
As it is well-known one of the most fascinating examples in
remarkable discussion between Einstein and Bohr on the conceptual
foundation of the quantum mechanics (Heisenberg energy-time
uncertainty relation especially) was an experimental device
representing a box hanged on an elastic spring. The pair of
similar devices is used in this work for formulation of a
classical and (implicitly) quantum mechanical analog of the famous
two capacitors paradox. It admits a simple solution of the paradox
since energy difference or seeming paradoxical "loss" can be
explained by work of the elastic force for moving of the boxes in
the gravitational field. (Obviously, original two capacitors
paradox can be explained in the analogous way.)
\end {abstract}

\vspace {1.5cm}

As it is well-known remarkable two-capacitors paradox, formulated
and considered in many textbooks and articles on the basic
principles and applications of the electronic and electrodynamics
[1]-[7], states the following. Consider an ideal (without any
electrical resistance and inductivity) electrical circuit with
first, initially charged, and second, initially non-charged, of
two identical capacitors. In given circuit, by transition from
initial, open state (switch OFF state) in the closed state (switch
ON state), an unexpected, mysterious loss of the half of initial
energy of electrical fields within capacitors occurs. Different
authors [4]-[7] suggest that given energy loss is realized by
different dissipative processes (Joule heating or/and
electromagnetic waves emissions) realized by non-neglectable
residual electric resistances and inductivities in realistic
circuits.

As it is well-known one of the most fascinating examples in
remarkable discussion between Einstein and Bohr on the conceptual
foundation of the quantum mechanics (Heisenberg energy-time
uncertainty relation especially) [8] was an experimental device
representing a box hanged on an elastic spring. The pair of
similar devices will be used in this work for formulation of a
classical and (implicitly) quantum mechanical analog of the two
capacitor paradox. It admits a simple solution of the paradox
since energy difference or seeming paradoxical "loss" can be
explained by work of the elastic force for moving of the box in
gravitational field. (Obviously, original two capacitors paradox
can be explained in the analogous way.)

In remarkable Einstein-Bohr discussion on the conceptual problems
of the quantum mechanics foundation (Heisenberg
coordinate-momentum and energy-time uncertainty relation
especially) [8] the following experimental device was especially
interesting. It represents a box, for example a cubic box, hanged
on a elastic spring in Earth gravitational field (practically
constant nearly Earth surface) so that elastic and gravitational
force are initially in enquilibrium. Given box holds an external
pointer that points out position of the box, i.e. equilibrium
point, on a vertical length scale fixed without box.

At the center of a box vertical side there is a small hole that
can be closed either open by corresponding mechanism connected
with a clock. When mechanism opens hole in a time moment
determined by clock single photon can leave the box. After photon
leave of the box mass of the box, according to equivalence
principle, becomes smaller and elastic force becomes larger than
gravitational. For this reason elastic force lift the box toward a
higher point on the scale representing new equilibrium point.
Given lifting can be considered as the work done by elastic force.

Now, we shall consider the pair of similar experimental devices
for formulation of a classical and quantum mechanical analog of
remarkable two capacitor paradox [1]-[7].

As well as in the mentioned Einstein-Bohr discussion we shall use
first box spring replelte completely by a liquid. Suppose that
total mass of the liquid initially equals M. Then equilibrium
condition between gravitational and elastic force $Mg=kX$, where k
represents spring elasticity coefficient and X - box position,
determines this position by expression
\begin {equation}
  X=\frac {Mg}{k}        .
\end {equation}
Energy of the elastic force in this position equals, as it is
well-known,
\begin {equation}
   E_{in 1}= \frac {kX^{2}}{2}= \frac {M^{2}g^{2}}{2k}
\end {equation}

After opening of the hole at vertical side of the box by mentioned
mechanism the following occurs. Through hole, in an admitable
approximation, discretely, drop by drop any of which holds mass
$m=\frac {M}{N}$ for $N\gg 1$, there is a free fall of the fluid
drops in the second, initially empty, neighbouring box, placed
immediately under the first box. This second box holds form
identical to the first, but it does not hold high horizontal side
so that free falling drops can arrive inside the second box.
Suppose, also, that given second box is placed at a vertical
spring as well as that this box holds a pointer which points out
position of the second box, i.e. equilibrium point between
gravitational force acting at liquid and elastic force. Since
second box is initially empty initial energy of corresponding
elastic force is zero.

In this way initial total energy of both elastic forces, elastic
force acting at the first box and elastic force acting at the
second box initially, equals
\begin {equation}
   E_{in} = E_{in 1}+ 0 = E_{in 1}
\end {equation}
that is identical to  $ E_{in 1}$ (2).

As it is not hard to see liquid will turn from the first in the
second box till final moment when masses in both boxes become
equivalent and equal $\frac {M}{2}$. In this moment energies of
the elastic force acting on the first and elastic force acting on
the second box will be equivalent and will equal
\begin {equation}
   E_{fin 1,2}= (\frac {M}{2})^{2}\frac {g^{2}}{2k} = \frac {1}{4}M^{2}\frac {g^{2}}{2k} = \frac {1}{4} E_{in 1}             .
\end {equation}

Then total energy of both elastic forces equals
\begin {equation}
   E_{fin}= 2 E_{fin 1,2}= \frac {1}{2} E_{in 1}= \frac {1}{2} E_{in}             .
\end {equation}

Obviously, final total energy of the elastic forces is two times
smaller that initial total energy of the elastic forces and we
have a (seemingly) paradoxical energy loss equivalent to one half
of given initial energy. This is, of course, a complete analogy
with two capacitor paradox [1]-[7].

For explanation of given seeming paradox consider dynamics of the
box and liquid, i.e. drops more detailedly.

Suppose firstly that distance between boxes are small and that
kinetic energy that drop obtains by gravitational force by free
faling between two boxes can be neglected.

After free falling of the first drop in the initially empty second
box mass increases for m till m. For this reason appears
gravitational force mg larger than zero elastic force of the
spring. It causes compression of the spring, i.e. moving of the
box down from 0 for $q=\frac {mg}{k}$ till q representing  new
equilibrium position. It corresponds to increase of the spring
elastic force energy from 0 for  $\frac {kq^{2}}{2}$ till $\frac
{kq^{2}}{2}$.

By simple induction we can conclude the following. After free
falling of the n-th drop in the second box with mass $(n-1)m$
before drop falling, mass of the box increases for m till nm. For
this reason new gravitational force nmg becomes larger than
elastic force of the spring $(n-1)kq$. It causes further
compression of the spring, i.e. moving of the box down from
$(n-1)q$ for $q=\frac {mg}{k}$ till na representing new
equilibrium position. It corresponds to increase of the spring
elastic force energy from $\frac {k(n-1)^{2}q^{2}}{2}$ till $\frac
{kn^{2}q^{2}}{2}$ . As it is not hard to see corresponding energy
difference can be expressed in the following way
\begin {equation}
   \Delta E_{2n}= \frac {kn^{2}q^{2}}{2}- \frac {k(n-1)^{2}q^{2}}{2}\simeq  knq^{2}=
   knq q = nmg q  \hspace{1cm} {\rm for} \hspace{0.5 cm} n\gg 1 .
\end {equation}
It can be considered as the work of the gravitational force by
moving of the mass nm for q. In this way increase of the energy of
elastic force represents here direct consequence of the positive
work of gravitational force.

Then, complete, positive, difference of the energy of elastic
force from initial 0 value till final (after free falling of
$\frac {N}{2}$ drops) value (2) can be simply obtained by formula
\begin {equation}
\Delta E_{2} = (1 + 2 + ... \frac {N}{2}) kq^{2} = \frac
{1}{2}\frac {N}{2}(1 + \frac {N}{2})kq^{2}\simeq \frac {1}{4}\frac
{1}{2}kN^{2}q^{2}= \frac {1}{4}M^{2}\frac {g^{2}}{2k}= E_{fin 2}.
\end {equation}
It is very important to be pointed out this positive difference of
the initial energy of elastic force is caused by positive work of
gravitational force at the second box with discretely chengeable
mass.

On the other side, after free falling of the first drop initially
second box mass $M=Nm$ decreases for m till $(N-1)m$. For this
reason new gravitational force $(N-1)mg$ becomes smaller than
elastic force of the spring $kNq$. It causes compression of the
spring, i.e. moving of the box up from $Nq$ for $q=\frac {mg}{k}$
till $(N-1)q$ representing  new equilibrium position (we consider
absolute value of the position!). It causes decrease of the spring
elastic force energy from $\frac {1}{2}kN^{2}q^{2}$ till $\frac
{1}{2}k(N-1)^{2}q^{2}\simeq \frac {1}{2}kN^{2}q^{2} - kNq^{2}$ for
energy difference $- kNq^{2}= - kNq q= - Nmg q $. As it is not
hard to see given energy difference can be considered as the
negative work of the gravitational force by moving of the mass Nm
for q.

By simple induction we can conclude the following. After free
falling of the n-th drop in the second box, mass of the first box
decreases from $(N-(n-1))m$  for m till $(N-n)m$. For this reason
new gravitational force $(N-n)mg$ becomes smaller than elastic
force of the spring $(N-(n-1))kq$. It causes further compression
of the spring, i.e. moving of the box up from $(N-(n-1))q$ for
$q=\frac {mg}{k}$ till $(N-n)q$ representing new equilibrium
position (we consider absolute value of the position!). It causes
decrease of the spring elastic force energy from
$k(N-(n-1))^{2}\frac {q^{2}}{2}$ till $k(N-n)^{2}\frac
{q^{2}}{2}$. As it is not hard to see corresponding energy
difference can be expressed in the following way
\begin {equation}
   \Delta E_{1n}= k(N-n)^{2}\frac {q^{2}}{2} - k(N-(n-1))^{2}\frac {q^{2}}{2}˜  - kn q^{2}=
   - knq q = - nmg q  \hspace{1cm} {\rm for} \hspace{0.5 cm} n \gg 1 .
\end {equation}
It can be considered as the negative work of the gravitational
force by moving of the mass nm for q. In this way decrease of the
energy of elastic force represents here direct consequence of the
negative work of gravitational force.

Then, complete, negative, difference of the energy of elastic
force from initial  value $\frac {1}{2}kN^{2}q^{2}$ till final
(after free falling of $\frac {N}{2}$ drops) value (2) can be
simply obtained by formula
\begin {equation}
   \Delta E_{1} = -(N + (N-1) + ... (N/2 - 1)) kq^{2} =
\end {equation}

\[-\frac {1}{2}\frac {N}{2}(1 + 3\frac {N}{2})kq^{2}\simeq -(\frac
{3}{4})\frac {1}{2}kN^{2}q^{2}= -(\frac {3}{4})M^{2}\frac
{g^{2}}{2k}= -(\frac {3}{4})E_{in}\].

 It is very important to be
pointed out this negative difference of the initial energy of
elastic force is caused by negative work of gravitational force at
the first box with discretely chengeable mass.

Total difference of the total energy of both elastic forces from
initial value (3) till final value (5) equals, according to (7)
and (9),
\begin {equation}
 \Delta E = \Delta E_{1} + \Delta E_{2} = = -\frac {3}{4}E_{in}+ \frac {1}{4}E_{in} =  - \frac {E_{in}}{2}    .
\end {equation}
In this way we obtain very simple and reasonable solution of two
box paradox. Simply speaking "loss", i.e. negative total
difference of the total elastic energy of both systems is result
of the total negative work of the gravitational force by moving of
the systems.

Finally, it can be observed that all this can be formulated
completely analogously in the though (gedanken) experiment form by
changing of the liquid drops by photons, i.e. by use of the pair
of original Einstein-Bohr devices. (Photon, of course, cannot
arise from the first in the other box by free fall. However,
photon can arise from the first in the other box by an appropriate
mirror.) It represents an interesting quantum analog of two
capacitor paradox, but it goes over basic intentions of given
work.

In conclusion, the following can be shortly repeated and pointed
out. As it is well-known one of the most fascinating examples in
remarkable discussion between Einstein and Bohr on the conceptual
foundation of the quantum mechanics (Heisenberg energy-time
uncertainty relation especially) was an experimental device
representing a box hanged on an elastic spring. The pair of
similar devices is used in this work for formulation of a
classical and (implicitly) quantum mechanical analog of the famous
two capacitor paradox. It admits a simple solution of the paradox
since energy difference or seeming paradoxical "loss" can be
explained by work of the elastic force for moving of the boxes in
the gravitational field.(Obviously, original two capacitors
paradox can be explained in the analogous way.)

\vspace{0.5cm}

Authors are deeply grateful to Prof. Dr. Tristan H$\ddot {\rm
u}$bsch for illuminating discussions.

\vspace{1.5cm}

 {\large \bf References}

\begin {itemize}

\item [[1]] D. Halliday, R. Resnick, {\it Physics, Vol. II} (J. Willey, New York, 1978)
\item [[2]] F. W. Sears, M.W. Zemansky, {\it University Physics} (Addison-Wesley, Reading, MA, 1964)
\item [[3]] M. A. Plonus, {\it Applied Electromagnetics}, (McGraw-Hill, New York, 1978)
\item [[4]] E. M. Purcell, {\it Electricity and Magnetism, Berkeley Physics Course Vol. II} (McGraw-Hill, New York, 1965)
\item [[5]] R. A. Powel, {\it Two-capacitor problem: A more realistic view}, Am. J. Phys. {\bf 47} (1979) 460
\item [[6]] T. B. Boykin, D. Hite, N. Singh, Am. J. Phys. {\bf 70} (2002) 460
\item [[7]] K. T. McDonald, {\it A Capacitor Paradox}, class-ph/0312031
\item [[8]] N. Bohr, {\it Atomic Physics and Human Knowledge} (John Wiley, New York , 1958)

\end {itemize}

\end {document}